\documentclass{article}
\usepackage[sglblindworkshop, final]{ai4nextg_neurips_2025}
\usepackage[utf8]{inputenc} 
\usepackage[T1]{fontenc}    
\usepackage{hyperref}  
\usepackage{url}
\usepackage{xurl}
\usepackage{booktabs}   
\usepackage{amsfonts}    
\usepackage{nicefrac}  
\usepackage{microtype}  
\usepackage{xcolor} 
\usepackage{graphicx}
\usepackage{svg}
\usepackage{comment}
\usepackage{bbding}
\usepackage{multirow}
\usepackage{longtable}
\usepackage{array}
\usepackage{amsmath}
\usepackage{amssymb}
\usepackage{indentfirst}
\usepackage{verbatim}
\usepackage{makecell}
\usepackage{soul}
\usepackage{tabularx}
\usepackage{subcaption}
\usepackage{lettrine}
\usepackage{algorithm}
\usepackage{algorithmic}
\usepackage{colortbl} 
\usepackage{subcaption}
\usepackage[shortcuts,acronym,nonumberlist]{glossaries}

\makeglossaries 
\newacronym{cnn}{CNN}{Convolutional Neural Network}
\newacronym{wpfm}{WPFM}{Wireless Physical Layer Foundation Model}
\newacronym{rf}{RF}{Radio Frequency}
\newacronym{ml}{ML}{Machine Learning}
\newacronym{NeSy}{NeSy}{Neuro-symbolic}

\title{Reasoning Meets Representation: Envisioning Neuro-Symbolic Wireless Foundation Models}

\author{%
  Jaron Fontaine\\
  Department of Information Technology\\
  Ghent University, Belgium\\
  \texttt{jaron.fontaine@ugent.be} \\
  \And
  Mohammad Cheraghinia\\
  Department of Information Technology\\
  Ghent University, Belgium\\
  \texttt{mohammad.cheraghinia@ugent.be} \\
  \And
  John Strassner\\
  Americas Standards and Industry Division\\
  Futurewei Technologies, USA\\
  \texttt{john.sc.strassner@futurewei.com} \\
 \And
  Adnan Shahid\\
  Department of Information Technology\\
  Ghent University, Belgium\\
  \texttt{adnan.shahid@ugent.be} \\
\And
  Eli De Poorter\\
  Department of Information Technology\\
  Ghent University, Belgium\\
  \texttt{eli.depoorter@ugent.be} \\
}

\begin{document}

\maketitle

\begin{abstract}
Recent advances in \glspl{wpfm} promise a new paradigm of universal \ac{rf} representations. However, these models inherit critical limitations found in deep learning such as the lack of explainability, robustness, adaptability, and verifiable compliance with physical and regulatory constraints. In addition, the vision for an AI-native 6G network demands a level of intelligence that is deeply embedded into the systems and is trustworthy. In this vision paper, we argue that the neuro-symbolic paradigm, which integrates data-driven neural networks with rule- and logic-based symbolic reasoning, is essential for bridging this gap. We envision a novel Neuro-Symbolic framework that integrates universal \ac{rf} embeddings with symbolic knowledge graphs and differentiable logic layers. This hybrid approach enables models to learn from large datasets while reasoning over explicit domain knowledge, enabling trustworthy, generalizable, and efficient wireless AI that can meet the demands of future networks. 
\end{abstract}

\section{Introduction}
\Ac{ml} for wireless systems has largely evolved along two paths: symbolic approaches grounded in domain knowledge, and sub-symbolic (data-driven, or neural) methods fueled by large-scale training. Symbolic methods offer transparency and explainability, which provides compliance with physical or regulatory constraints, yet struggle with addressing high levels of complexity and adaptability. Neural methods excel at learning complex patterns from raw signal data but typically function as "black boxes," limiting their reliability and trustworthiness.

Recently, large foundation models pre-trained on diverse \acf{rf} data are proposed to create universal representations \citep{10615509, guo2025largeaimodelswireless}, which is a significant step forward. Industry initiatives are already aiming for \acfp{wpfm} to achieve "semantic \ac{rf} understanding" \citet{fontaine2023reimagining} and can transfer knowledge across tasks such as modulation recognition, interference classification, sensing, resource management, and localization. However, such neural network foundation models still struggle with addressing trust and explainability. 

We argue that a neuro-symbolic approach, which integrates data-driven neural networks with rule- and logic-based symbolic reasoning, is crucial to build the next generation of wireless foundation models \citet{10682967}. 
In this paper, we propose the principles of a \ac{NeSy} AI framework, which combines the perceptual power of deep learning with formal reasoning, to the unique constraints and opportunities of the wireless physical layer. By integrating the flexibility of neural networks with the explainability and formal guarantees of symbolic systems, we can create foundation models that are not only powerful but also trustworthy. The key motivations and the distinct advantages of our proposed solution over existing approaches are summarized in Table \ref{tab:comparison_metrics}.







\begin{table*}[h]
\centering
\small
\renewcommand{\arraystretch}{1.3}
\setlength{\tabcolsep}{6pt}
\begin{tabular}{
    >{\raggedright\arraybackslash}m{2.2cm} 
    >{\raggedright\arraybackslash}m{3.2cm} 
    >{\raggedright\arraybackslash}m{3.8cm} 
    >{\raggedright\arraybackslash}m{3.0cm}
}
\toprule
\textbf{Metric} & 
\makecell{\textbf{Symbolic Methods}\\(\citealp{bhuyan2024neuro})} &
\makecell{\textbf{Neural Methods}\\(\citealp{schwalbe2024comprehensive})} &
\makecell{\textbf{Neuro-symbolic}\\\textbf{methods with}\\\textbf{WPFM interaction}\\(Our vision)} \\
\midrule

\textit{Explainability} & High (explicit rules \& formulae) & Low (“black box”) & High (symbolic reasoning) \\

\textit{Scalability} & Low (brittle; rule explosion, complex logic) & High (learns from data) & High (leverages neural scalability) \\

\textit{Adaptability} & Low (requires manual editing) & High (via ICL, etc.) & Very High (reasons over knowledge + few-shot) \\

\textit{Constraint Compliance} & High (hard-coded) & Low (unable to prove) & High (verifiable via symbolic proofs) \\

\textit{Data Efficiency} & High (uses domain knowledge) & Low (needs massive datasets) & Medium (priors reduce data needs) \\
\bottomrule
\end{tabular}
\caption{Comparison of Symbolic, Neural, and Neuro-symbolic (WPFM-integrated) methods across key metrics.}
\label{tab:comparison_metrics}
\end{table*}

\section{A Neuro-Symbolic Framework for Wireless AI}
\begin{figure}
    \centering
    \includegraphics[width=\linewidth]{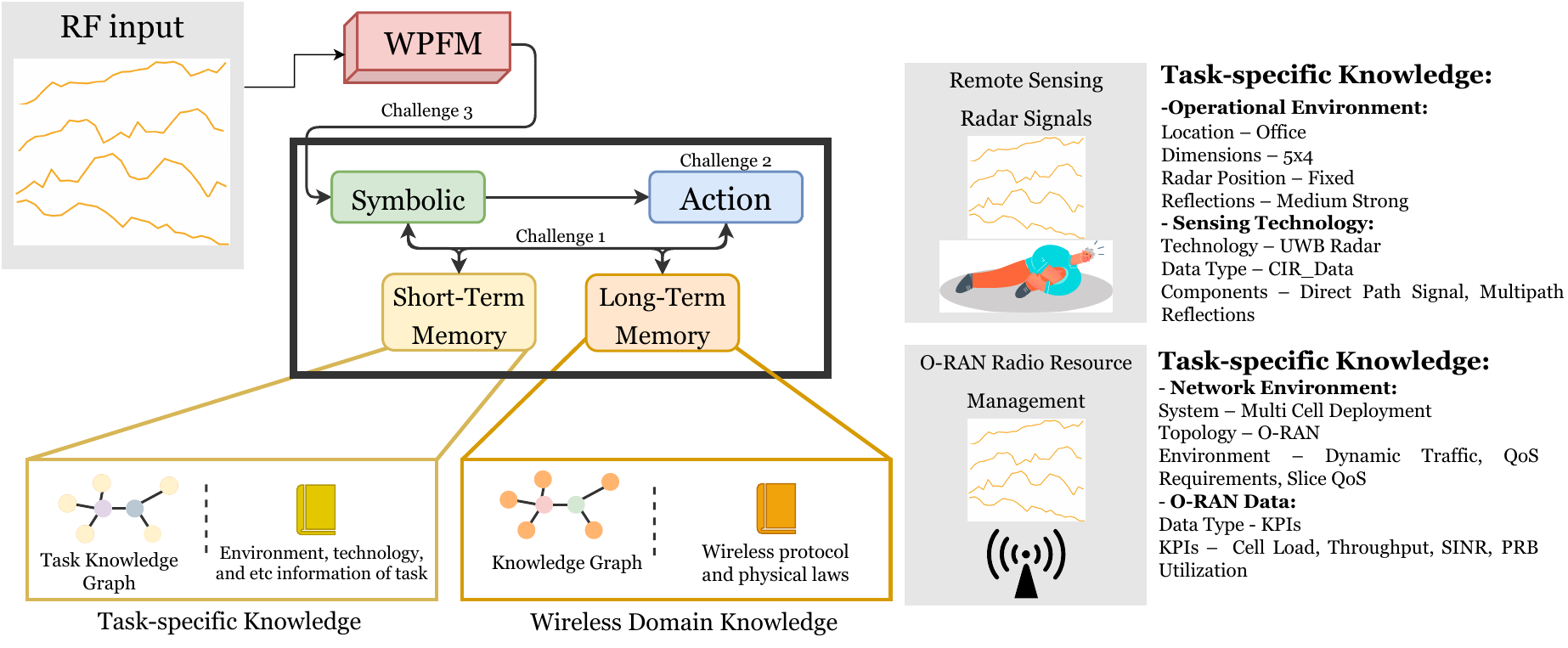}
    \caption{A Neuro-Symbolic Framework for Wireless AI: that comprises of a WPFM that embeds RF signals to be used by the symbolic component. The symbolic component utilizes memories, which include task-specific and general information from the wireless domain, to generate interpretable symbols. Using symbolic reasoning, actions can be performed, e.g., fall detection or O-RAN radio resource management, which are included as examples. In the right part of this figure, we included task-specific knowledge examples on top of general wireless domain knowledge, which aim to improve the accuracy and explainability of the system simultaneously. }
    \label{fig:diagram}
\end{figure}

In this section, we discuss the proposed framework for wireless problems utilizing \ac{wpfm} and neuro-symbolic methods which is illustrated in Figure~\ref{fig:diagram}.

\subsection{WPFM: The Neural Perception Engine}
The field of wireless communications is struggling with an unsustainable paradigm of training specialized models for each specific task. Focusing on one task at a time is expensive and slow, from gathering large datasets to repetitive deployment. Recent advances in \ac{rf} foundation models demonstrate the feasibility of universal representations trained on diverse datasets \citep{cheraghinia2025foundationmodelwirelesstechnology, zhou2025spectrumfmfoundationmodelintelligent}.

The pre-training of such foundation models, typically Transformers, does not rely on extensively labeled datasets but can learn from unlabeled data in a self-supervised way. To do so, masking or next-sample prediction can be applied to learn rich physical-layer embeddings. The input data of the \ac{wpfm} can be represented as raw time series, e.g. in-phase quadrature (IQ) samples, channel impulse response (CIR), channel state information (CSI), while tokenization can be performed using patching. To retain the position of the patches during the self-attention mechanism, positional or learned encoding schemes are applied. After the \ac{wpfm} is finetuned on a large variety of RF data, a lightweight finetuning process helps the model to quickly adapt to new downstream tasks (classification or predictions) with minimal labeling requirements.

In this paper, we propose that these foundation models can be further improved by producing embeddings (sub-symbolics), instead of traditional label classifications or predictions, anticipating the need for interpretable decision-making and reusable information for different tasks from a single model. These foundation models act as the "eye" of the process, receiving \ac{rf} modalities and producing input to the symbolic module.

\subsection{A Modular Neuro-Symbolic Wireless AI Architecture}
To overcome the limitations of a purely neural approach, we propose extending the \ac{wpfm} within a modular neuro-symbolic architecture, adopting a neuro-symbolic pipeline \cite{10682967, math13111707}. This architecture cleanly separates the tasks of perception and reasoning. The neural perception transforms raw \ac{rf} signals into a dense embedding. Next, the symbolic module integrates these embeddings with task-specific and wireless domain knowledge. This modular design grounds the neural network's pattern recognition in verifiable logic, enabling explainability and constraint enforcement.

The (neuro) symbolic reasoning pipeline comprises of three components: (1) ontologies to provide a shared semantic schema, (2) a knowledge graph, which is a concrete instance of the ontologies, to represent a structured world model, and (3) differentiable logic to make the symbolic reasoning process itself trainable via gradient-based methods. By representing logical rules as continuous functions, it becomes possible to backpropagate gradients from the symbolic layer through to the neural \ac{wpfm}, enabling end-to-end training of the entire hybrid system. This transformation is often accomplished using techniques from fuzzy logic, where discrete operators like AND and OR are replaced by continuous, differentiable functions known as t-norms \cite{van2022analyzing}. 

\section{Research Challenges \& Open Questions}
The integration of symbolic reasoning with large-scale neural embeddings for wireless systems presents several unique research challenges.

\textbf{Challenge 1: Knowledge Representation for the Physical Layer.} A fundamental challenge is in creating a formal, machine-readable knowledge base for the wireless domain. It is not enough to just digitize standards documents; it requires encoding the continuous laws of physics (e.g., Maxwell's equations, channel fading models) and the discrete logic of communication protocols (e.g., MAC layer state transitions) into a unified formalism. This challenge needs to answer if a wireless knowledge graph can represent both continuous and discrete logics in a way that is compliant with robust symbolic reasoning.

\textbf{Challenge 2: Real-Time Neuro-Symbolic Inference.} Wireless physical layer systems operate under strict, microsecond-level constraints. While \acp{wpfm} are computationally intensive, their inference latency can be improved using model compression techniques such as knowledge distillation to train smaller, faster models, quantization to reduce numerical precision for hardware acceleration, and pruning to eliminate redundant parameters. However, symbolic reasoning can also be slow. As such, it becomes a major challenge to achieve real-time inference with the entire hybrid model. This leads to the question whether we can leverage techniques like differentiable logic circuits, which can be compiled directly to hardware (FPGAs/ASICs), to create a reasoning layer that meets the demanding deadlines of physical layer control loops.

\textbf{Challenge 3: Bridging the Sub-symbolic (embeddings) to Symbolic Gap.} A core challenge in all neuro-symbolic AI is the inherent mismatch in knowledge representation. The self-attention mechanisms of foundation models create distributed, non-local representations (sub-symbolic) that have complex interactions. In contrast, symbolic AI represents knowledge explicitly using discrete, localized structures like symbols and rules. To overcome this challenge, the gap between the continuous, distributed representations of neural networks and the discrete, structured representations of symbolic logic needs to be bridged. Finding semantically meaningful and computationally effective ways to translate information across this gap is a research problem motivating this work.

This vision opens serveral critical questions to be addressed the research community. First, what is the inherent trade-off between the expressive richness of the symbolic representation and the real-time inference capability of a \ac{NeSy}-\ac{wpfm}? Does a Pareto frontier exist between reasoning depth and latency in this domain? Second, beyond task-specific accuracy, how do we evaluate the "trustworthiness" of a \ac{NeSy}-\ac{wpfm}? What novel benchmarks and methodologies are required to quantify improvements in interpretability and robustness to out-of-distribution scenarios, such as novel interference types or zero-day exploits at the physical layer?

\section{Impact on Applications}
The development of neuro-symbolic \acp{wpfm} will be a foundational step toward realizing the vision of AI-native 6G \citet{10273257}. This vision is about designing networks where AI is "natively integrated from the ground up" into the core fabric of the system. A \ac{NeSy}-\ac{wpfm} solution represents a concrete instantiation of this principle at the physical layer, providing the Radio Access Network with the capability for reasoning, self-explanation, and verifiable compliance.

This framework lays the foundation for applications such as:

\textbf{Trustworthy and Verifiable Wireless AI:} Enabling systems that can explain their decisions (e.g., "I am allocating this resource block because of predicted interference from source X, which violates regulatory mask Y"), generalize to unseen network conditions, and guarantee compliance to physical laws and spectrum policies.

\textbf{Cognitive and Autonomous 6G RAN:} Embedding reasoning directly into physical-layer intelligence, moving from today's reactive optimization to proactive, self-evolving networks that can autonomously manage resources, mitigate interference, and diagnose faults with minimal human intervention. 

\textbf{Unified Sensing and Communication:} Integrating sensing and communication under a single, reasoning-aware solution unlocks new cross-domain applications, such as remote sensing in healthcare, localization for radio resource management.


\section{Conclusion and future work}
By combining universal neural \ac{rf} representations with symbolic reasoning, we can overcome the limitations of purely data-driven approaches and move toward trustworthy, generalizable, and efficient systems. This hybrid approach is essential for realizing the vision of an AI-native wireless system where intelligence is robustly embedded into the system.

A primary direction for future work is to investigate these neuro-symbolic capabilities to empower agentic AI. The reasoning offered by this framework can serve as the cognitive engine for autonomous agents capable of network management, self-diagnosis, and optimization with minimal human intervention. However, a requirement for such advanced agents is a standardized, machine-readable knowledge base. Therefore, another future direction can be developing a standardized, open-source wireless knowledge graph. This involves encoding the continuous laws of physics and the discrete logic of communication protocols into a unified structure, providing the necessary information for robust symbolic reasoning.

\bibliographystyle{plainnat}
\bibliography{references}

\end{document}